\documentclass{article}

\usepackage[utf8]{inputenc} 
\usepackage{setspace} 
\usepackage{cite} 
\usepackage{graphicx} 
\usepackage{braket} 
\usepackage{amsmath} 
\usepackage[margin=1in]{geometry} 
\usepackage{gensymb} 
\usepackage{hyperref}
\usepackage{graphicx}
\usepackage[usenames, dvipsnames]{color}
\usepackage{sectsty}
\usepackage{esvect} 
\usepackage{mwe}
\usepackage{subfig} 
\usepackage{amsfonts} 
\usepackage{bbold} 
\usepackage{lscape}
\usepackage{authblk}
\sectionfont{\fontsize{12}{15}\selectfont}
\subsectionfont{\fontsize{10}{12}\selectfont}
\title{Comment on `Supervised quantum machine learning models are kernel methods'}
\author[1,2]{Rajiv Krishnakumar}
\affil[1]{QuantumBasel, Schorenweg 44b, 4144 Arlesheim, Switzerland}
\affil[2]{Center for Quantum Computing and Quantum Coherence (QC2),
University of Basel, Klingelbergstrasse 82, 4056 Basel, Switzerland}
\date{}

\begin{document}
\maketitle
\begin{abstract}
    We identify a few small errors in the proof of Theorem 1 in M. Schuld, `Supervised quantum machine learning models are kernel methods'\cite{maria}, Appendix A. These do not affect the validity of the theorem, but do matter if one wants to use the appendix to explicitly compute the Fourier coefficients of a quantum kernel. We give the corrected derivation, and additionally show that the paper's worked cosine-kernel example contains a second, unrelated error that happens to cancel the first, thereby coincidentally yielding the correct result.
\end{abstract}
\setlength{\parindent}{0pt} 

\section{Summary of note}
\label{sec:summary}
We make a small examination of a proof of Theorem 1 in \cite{maria}. This theorem states that for a time-evolution encoding embedding of the form
\begin{align}
	S(\boldsymbol{x}) = W^{(N+1)}e^{-ix_NG}W^{(N)}\ldots W^{(2)}e^{-ix_1G}W^{(1)}
	\label{eq:circuit}
\end{align}

where $W^{(i)}$ are arbitrary unitary operators and $G$ are generating Hamiltonians, the corresponding quantum kernel can be written as

\begin{align}
	k(\boldsymbol{x},\boldsymbol{x'}) = \sum_{\boldsymbol{s},\boldsymbol{t} \in \boldsymbol{\Omega}}e^{-i\boldsymbol{s}\boldsymbol{x}}e^{i\boldsymbol{t}\boldsymbol{x'}}c_{\boldsymbol{s}\boldsymbol{t}}
\end{align}

where $\boldsymbol{\Omega} \subseteq \mathbb{R}^N$ and $c_{\boldsymbol{s}\boldsymbol{t}} \subseteq \mathbb{C}$ where for every $\boldsymbol{s},\boldsymbol{t} \in \boldsymbol{\Omega}$, we have $c_{-\boldsymbol{s}-\boldsymbol{t}}=c^*_{\boldsymbol{s}\boldsymbol{t}}$.\\
\\
In Appendix 1 of \cite{maria}, the author proceeds to prove this theorem and elaborate on the details. It is in this section that we seem to have found a couple of errors. This does not change the validity of the theorem, but could make a difference if one wants to use the details in the appendix to explicitly compute the Fourier representation of a kernel. The three errors are
\begin{enumerate}
	\item In Equation (A5), the terms $W_{1k_1}^{(1)}\ldots W_{k_{N-1}k_N}^{(N)}$ should have the indices reversed and therefore actually be $W_{k_11}^{(1)}\ldots W_{k_Nk_{N-1}}^{(N)}$. This comes from the fact that when taking the conjugate transpose of $W^{(i)}$ and multiplying it from the left by a vector $\boldsymbol{v}^T$, where $\boldsymbol{v} \in \mathbb{R}^d$, the resulting vector is $\left(\sum_{j=1}^dv_jW^{(i)*}_{1j},\ldots,\sum_{j=1}^dv_jW^{(i)*}_{dj}\right)$.
	\item In Equation (A5), the expression should be multiplied by the extra term $\delta_{j_N,k_N}$. This comes from the fact that when taking the expectation of the multiplication of matrices, there should be an overlapping term in the summation. This has the further consequences that in Equation (A6) there should also be the same extra term and in Equation (A7) and (A8), the expressions should be multiplied by $\delta_{j_N,k_N}\delta_{h_N,l_N}$. 
	\item In Equation (A7) and (A8), $w^*_{\boldsymbol{h}}$ should be $w_{\boldsymbol{h}}$ and $w_{\boldsymbol{l}}$ should be $w^*_{\boldsymbol{l}}$. This is because when taking the absolute value squared, the terms $W_{l_il_{i-1}}$ are conjugate and similary with the terms $W^*_{h_ih_{i-1}}$, and the so final terms that are conjugated are swtiched. 
\end{enumerate}
We elaborate on details of how we came to these three points in the section below.

\section{Derivation of the Fourier representation of quantum kernels}
We will now derive the explicit expression for a quantum kernel given \autoref{eq:circuit} and show how we arrive to the final results with the corrected errors mentioned in \autoref{sec:summary}. Like in \cite{maria}, we will assume that $e^{-ix_iG}$ can be written as $V^{(i)}e^{-ix_i\Sigma}V^{(i)\dagger}$ where
\begin{align}
	\Sigma &= 	\begin{pmatrix}
		\lambda_1 & 0 & \ldots\\ 
		0 & \lambda_2  \\ 
		\vdots  & & \ddots\\
		& & & \lambda_d
	\end{pmatrix}
\end{align}
and $V^{(i)}$ and $V^{(i)\dagger}$ are arbitrary unitary matrices that get absorbed into the respective $W^{(i)}$ matrices next to them, and we we will redefine these combined matrices as $W^{(i)}$. OK, here we go!
\begin{landscape}
\scriptsize
\begin{align}
	&\bra{0}S(\boldsymbol{x'})S(\boldsymbol{x})\ket{0} \\
	&= \bra{0} W^{(1)\dagger} \left(e^{-ix_1'\Sigma}\right)^\dagger W^{(2)\dagger}\ldots W^{(N)\dagger}\left(e^{-ix_N'\Sigma}\right)^\dagger\underbrace{W^{(N+1)\dagger}W^{(N+1)}}_{\mathbb{1}} e^{-ix_N\Sigma}W^{(N)} \ldots W^{(2)}e^{-ix_1\Sigma}W^{(1)}\ket{0} \\
&= \bra{0} W^{(1)\dagger} \left(e^{-ix_1'\Sigma}\right)^\dagger W^{(2)\dagger}\ldots W^{(N)\dagger}\left(e^{-ix_N'\Sigma}\right)^\dagger e^{-ix_N\Sigma}W^{(N)} \ldots W^{(2)}e^{-ix_1\Sigma}W^{(1)}\ket{0} \\
	&= \begin{pmatrix}
		1 \\ 0 \\ \vdots \\ 0
	\end{pmatrix}^T
	\begin{pmatrix}
		W^{(1)*}_{11} & W^{(1)*}_{21} & \ldots & W^{(1)*}_{d1} \\  
		W^{(1)*}_{12} & \ddots & & \vdots \\
		\vdots \\
		W^{(1)*}_{1d} & \ldots &  &W^{(1)*}_{dd}
	\end{pmatrix}
	\left(e^{-ix_1'\Sigma}\right)^\dagger W^{(2)\dagger}\ldots W^{(N)\dagger}\left(e^{-ix_N'\Sigma}\right)^\dagger e^{-ix_N\Sigma}W^{(N)} \ldots W^{(2)}
	\begin{pmatrix}
		W^{(1)}_{11} & W^{(1)}_{12} & \ldots & W^{(1)}_{1d} \\  
		W^{(1)}_{21} & \ddots & & \vdots \\
		\vdots \\
		W^{(1)}_{d1} & \ldots &  &W^{(1)}_{dd}
	\end{pmatrix}
	\begin{pmatrix}
		1 \\ 0 \\ \vdots \\ 0
	\end{pmatrix} \\
	&= \begin{pmatrix}
		W^{(1)*}_{11} \\ W^{(1)*}_{21} \\ \vdots \\ W^{(1)*}_{d1}
	\end{pmatrix}^T
	\left(e^{-ix_1'\Sigma}\right)^\dagger W^{(2)\dagger}\ldots W^{(N)\dagger}
	\begin{pmatrix}
		e^{ix_N'\lambda_1} & 0 & \ldots\\ 
		0 & e^{ix_N'\lambda_2}  \\ 
		\vdots  & & \ddots\\
		& & & e^{ix_N'\lambda_d}
	\end{pmatrix}
	\begin{pmatrix}
		e^{-ix_N\lambda_1} & 0 & \ldots\\ 
		0 & e^{-ix_N\lambda_2}  \\ 
		\vdots  & & \ddots\\
		& & & e^{-ix_N\lambda_d}
	\end{pmatrix}
	W^{(N)} \ldots W^{(2)}e^{-ix_1\Sigma}
	\begin{pmatrix}
		W^{(1)}_{11} \\ W^{(1)}_{21} \\ \vdots \\ W^{(1)}_{d1}
	\end{pmatrix} \\
	&= \begin{pmatrix}
		W^{(1)*}_{11}  e^{ix_1'\lambda_1} \\ W^{(1)*}_{21} e^{ix_1'\lambda_2} \\ \vdots \\ W^{(1)*}_{d1} e^{ix_1'\lambda_d}
	\end{pmatrix}^T
	W^{(2)\dagger}\ldots W^{(N)\dagger}
	\begin{pmatrix}
		e^{-i(x_N-x_N')\lambda_1} & 0 & \ldots\\ 
		0 & e^{-i(x_N-x_N')\lambda_2}  \\ 
		\vdots  & & \ddots\\
		& & & e^{-i(x_N-x_N')\lambda_d}
	\end{pmatrix}
	W^{(N)} \ldots W^{(2)}
	\begin{pmatrix}
		e^{-ix_1\lambda_1} W^{(1)}_{11} \\ e^{-ix_1\lambda_2} W^{(1)}_{21} \\ \vdots \\ e^{-ix_1\lambda_d} W^{(1)}_{d1}
	\end{pmatrix} \\
	&= \begin{pmatrix}
		\sum_{k_1=1}^d W^{(1)*}_{k_11}  e^{ix_1'\lambda_{k_1}} W^{(2)*}_{1k_1}\\ \vdots \\ \sum_{k_1=1}^d W^{(1)*}_{k_11}  e^{ix_1'\lambda_{k_1}} W^{(2)*}_{dk_1}
	\end{pmatrix}^T
	\left(e^{-ix_2'\Sigma}\right)^\dagger \ldots W^{(N)\dagger}
	\begin{pmatrix}
		e^{-i(x_N-x_N')\lambda_1} \\ 
		 & \ddots  \\ 
		& &  e^{-i(x_N-x_N')\lambda_d}
	\end{pmatrix}
	W^{(N)} \ldots e^{-ix_2\Sigma}
	\begin{pmatrix}
		\sum_{j_1=1}^d W^{(2)}_{1j_1} e^{-ix_1\lambda_{j_1}} W^{(1)}_{j_11} \\ \vdots \\ \sum_{j_1=1}^d W^{(2)}_{1j_d} e^{-ix_1\lambda_{j_1}} W^{(1)}_{j_11}
	\end{pmatrix} \\
		&= \begin{pmatrix}
		\sum_{k_1=1}^d W^{(1)*}_{k_11}  e^{ix_1'\lambda_{k_1}} W^{(2)*}_{1k_1} e^{ix_2'\lambda_{1}}\\ \vdots \\ \sum_{k_1=1}^d W^{(1)*}_{k_11}  e^{ix_1'\lambda_{k_1}} W^{(2)*}_{dk_1} e^{ix_2'\lambda_{d}}
	\end{pmatrix}^T
	 W^{(3)\dagger} \ldots W^{(N)\dagger}
	\begin{pmatrix}
		e^{-i(x_N-x_N')\lambda_1} \\ 
		 & \ddots  \\ 
		& &  e^{-i(x_N-x_N')\lambda_d}
	\end{pmatrix}
	W^{(N)} \ldots W^{(3)}
	\begin{pmatrix}
		e^{-ix_2\lambda_{1}}\sum_{j_1=1}^d W^{(2)}_{1j_1} e^{-ix_1\lambda_{j_1}} W^{(1)}_{j_11} \\ \vdots \\ e^{-ix_2\lambda_{d}} \sum_{j_1=1}^d W^{(2)}_{1j_d} e^{-ix_1\lambda_{j_1}} W^{(1)}_{j_11}
	\end{pmatrix} \\
	&= \begin{pmatrix}
		\sum_{k_1, k_2=1}^d W^{(1)*}_{k_11}  e^{ix_1'\lambda_{k_1}} W^{(2)*}_{k_2k_1} e^{ix_2'\lambda_{k_2}}W^{(3)*}_{1k_2}\\ \vdots \\ \sum_{k_1, k_2=1}^d W^{(1)*}_{k_11}  e^{ix_1'\lambda_{k_1}} W^{(2)*}_{k_2k_1} e^{ix_2'\lambda_{k_2}}W^{(3)*}_{dk_2}
	\end{pmatrix}^T
	\left(e^{-ix_3'\Sigma}\right)^\dagger \ldots W^{(N)\dagger}
	\begin{pmatrix}
		e^{-i(x_N-x_N')\lambda_1} \\ 
		 & \ddots  \\ 
		& &  e^{-i(x_N-x_N')\lambda_d}
	\end{pmatrix}
	W^{(N)} \ldots e^{-ix_3\Sigma}
	\begin{pmatrix}
		\sum_{j_1,j_2=1}^d W^{(3)}_{1j_2} e^{-ix_2\lambda_{j_2}} W^{(2)}_{j_2j_1} e^{-ix_1\lambda_{j_1}} W^{(1)}_{j_11} \\ \vdots \\ \sum_{j_1, j_2=1}^d W^{(3)}_{dj_2} e^{-ix_2\lambda_{j_2}} W^{(2)}_{1j_d} e^{-ix_1\lambda_{j_1}} W^{(1)}_{j_11}
	\end{pmatrix} \\
	&\vdots \\
	&= \begin{pmatrix}
		\sum_{k_1,\ldots, k_{N-1}=1}^d W^{(1)*}_{k_11}  e^{ix_1'\lambda_{k_1}} W^{(2)*}_{k_2k_1} e^{ix_2'\lambda_{k_2}}\ldots e^{ix_{N-1}'\lambda_{k_{N-1}}} W^{(N)*}_{1k_{N-1}}\\ \vdots \\ \sum_{k_1,\ldots, k_{N-1}=1}^d W^{(1)*}_{k_11}  e^{ix_1'\lambda_{k_1}} W^{(2)*}_{k_2k_1} e^{ix_2'\lambda_{k_2}}\ldots e^{ix_{N-1}'\lambda_{k_{N-1}}} W^{(N)*}_{dk_{N-1}}
	\end{pmatrix}^T
	\begin{pmatrix}
		e^{-i(x_N-x_N')\lambda_1} \\ 
		 & \ddots  \\ 
		& &  e^{-i(x_N-x_N')\lambda_d}
	\end{pmatrix}
	\begin{pmatrix}
		\sum_{j_1,\ldots,j_{N-1}=1}^d W^{(N)}_{1j_{N-1}} e^{-ix_{N-1}\lambda_{j_{N-1}}}\ldots W^{(2)}_{j_2j_1} e^{-ix_1\lambda_{j_1}} W^{(1)}_{j_11} \\ \vdots \\ \sum_{j_1,\ldots,j_{N-1}=1}^d W^{(N)}_{dj_{N-1}} e^{-ix_{N-1}\lambda_{j_{N-1}}}\ldots W^{(2)}_{j_2j_1} e^{-ix_1\lambda_{j_1}} W^{(1)}_{j_11}
	\end{pmatrix}
\end{align}
\begin{align}
	&= \begin{pmatrix}
		\sum_{k_1,\ldots, k_{N-1}=1}^d W^{(1)*}_{k_11}  e^{ix_1'\lambda_{k_1}} W^{(2)*}_{k_2k_1} e^{ix_2'\lambda_{k_2}}\ldots e^{ix_{N-1}'\lambda_{k_{N-1}}} W^{(N)*}_{1k_{N-1}}\\ \vdots \\ \sum_{k_1,\ldots, k_{N-1}=1}^d W^{(1)*}_{k_11}  e^{ix_1'\lambda_{k_1}} W^{(2)*}_{k_2k_1} e^{ix_2'\lambda_{k_2}}\ldots e^{ix_{N-1}'\lambda_{k_{N-1}}} W^{(N)*}_{dk_{N-1}}
	\end{pmatrix}^T
	\begin{pmatrix}
		e^{-i(x_N-x_N')\lambda_1}\sum_{j_1,\ldots,j_{N-1}=1}^d W^{(N)}_{1j_{N-1}} e^{-ix_{N-1}\lambda_{j_{N-1}}}\ldots W^{(2)}_{j_2j_1} e^{-ix_1\lambda_{j_1}} W^{(1)}_{j_11} \\ \vdots \\ e^{-i(x_N-x_N')\lambda_d}\sum_{j_1,\ldots,j_{N-1}=1}^d W^{(N)}_{dj_{N-1}} e^{-ix_{N-1}\lambda_{j_{N-1}}}\ldots W^{(2)}_{j_2j_1} e^{-ix_1\lambda_{j_1}} W^{(1)}_{j_11}
	\end{pmatrix} \\
	&= \sum_{k_1,\ldots, k_{N-1}=1}^d W^{(1)*}_{k_11} W^{(2)*}_{k_2k_1}... W^{(N-1)*}_{k_{N-1}k_{N-2}}  e^{i(x_1'\lambda_{k_1}-+x_2'\lambda_{k_2}+\ldots+x_{N-1}'\lambda_{k_{N-1}})} \sum_{j_1,\ldots,j_{N-1}=1}^d W^{(1)}_{j_11} W^{(2)}_{j_2j_1}... W^{(N-1)}_{j_{N-1}j_{N-2}} e^{-i(x_1\lambda_{j_1}+x_2\lambda_{j_2}+\ldots+x_{N-1}\lambda_{j_{N-1}})} \times e^{-i(x_N-x_N')\lambda_1} W^{(N)*}_{1k_{N-1}} W^{(N)}_{1j_{N-1}}\\
	&+ \ldots + \\
	&=\sum_{k_1,\ldots, k_{N-1}=1}^d W^{(1)*}_{k_11} W^{(2)*}_{k_2k_1}... W^{(N-1)*}_{k_{N-1}k_{N-2}}  e^{i(x_1'\lambda_{k_1}-+x_2'\lambda_{k_2}+\ldots+x_{N-1}'\lambda_{k_{N-1}})} \sum_{j_1,\ldots,j_{N-1}=1}^d W^{(1)}_{j_11} W^{(2)}_{j_2j_1}... W^{(N-1)}_{j_{N-1}j_{N-2}} e^{-i(x_1\lambda_{j_1}+x_2\lambda_{j_2}+\ldots+x_{N-1}\lambda_{j_{N-1}})} \times  e^{-i(x_N-x_N')\lambda_d} W^{(N)*}_{dk_{N-1}} W^{(N)}_{dj_{N-1}}\\
	&=\sum_{k_1,\ldots, k_{N-1}=1}^d W^{(1)*}_{k_11} W^{(2)*}_{k_2k_1}... W^{(N-1)*}_{k_{N-1}k_{N-2}}  e^{i(x_1'\lambda_{k_1}-+x_2'\lambda_{k_2}+\ldots+x_{N-1}'\lambda_{k_{N-1}})} \sum_{j_1,\ldots,j_{N-1}=1}^d W^{(1)}_{j_11} W^{(2)}_{j_2j_1}... W^{(N-1)}_{j_{N-1}j_{N-2}} e^{-i(x_1\lambda_{j_1}+x_2\lambda_{j_2}+\ldots+x_{N-1}\lambda_{j_{N-1}})} \times \sum_{k_N=1}^d e^{-i(x_N-x_N')\lambda_{k_N}}W^{(N)*}_{k_Nk_{N-1}} W^{(N)}_{k_Nj_{N-1}}\\
	\label{eq:full}
	&= \sum_{j_1,\ldots, j_N=1}\sum_{k_1,\ldots, k_N=1}e^{-i(x_1\lambda_{j_1}-x'_1\lambda_{k_1}+\ldots+x_N\lambda_{j_N}-x'_N\lambda_{k_N})}W^{(1)*}_{k_11}\ldots W^{(N)*}_{k_Nk_{N-1}}W^{(1)}_{j_11}\ldots W^{(N)}_{j_Nj_{N-1}}\delta_{j_Nk_N} \\
	&= \sum_{\boldsymbol{j}}\sum_{\boldsymbol{k}}e^{-i(\boldsymbol{\Lambda_j}\boldsymbol{x}-\boldsymbol{\Lambda_k}\boldsymbol{x'})}w^*_{\boldsymbol{k}}w_{\boldsymbol{j}}\delta_{j_Nk_N}
\end{align}
\end{landscape}

\normalsize
We now use this to compute the kernel
\begin{align}
	k(\boldsymbol{x},\boldsymbol{x'}) &= |\bra{0}S(\boldsymbol{x'})S(\boldsymbol{x})\ket{0}|^2 \\
	&= \left|\sum_{\boldsymbol{j}}\sum_{\boldsymbol{k}}e^{-i(\boldsymbol{\Lambda_j}\boldsymbol{x}-\boldsymbol{\Lambda_k}\boldsymbol{x'})}w^*_{\boldsymbol{k}}w_{\boldsymbol{j}}\delta_{j_Nk_N}\right|^2\\
	&= \sum_{\boldsymbol{j}}\sum_{\boldsymbol{k}}\sum_{\boldsymbol{h}}\sum_{\boldsymbol{l}}e^{-i(\boldsymbol{\Lambda_j}-\boldsymbol{\Lambda_l})\boldsymbol{x})}e^{i(\boldsymbol{\Lambda_k}-\boldsymbol{\Lambda_h})\boldsymbol{x'})}w^*_{\boldsymbol{k}}w^*_{\boldsymbol{l}}w_{\boldsymbol{j}}w_{\boldsymbol{h}}\delta_{j_Nk_N}\delta_{l_Nh_N} \\
	&= \sum_{\boldsymbol{s},\boldsymbol{t} \in \boldsymbol{\Omega}}e^{-i\boldsymbol{s}\boldsymbol{x}}e^{i\boldsymbol{t}\boldsymbol{x'}} \sum_{\boldsymbol{j},\boldsymbol{l}|\boldsymbol{\Lambda_j}-\boldsymbol{\Lambda_l}=\boldsymbol{s}}\sum_{\boldsymbol{k},\boldsymbol{h}|\boldsymbol{\Lambda_k}-\boldsymbol{\Lambda_h}=\boldsymbol{t}}w_{\boldsymbol{j}}w_{\boldsymbol{h}}w^*_{\boldsymbol{k}}w^*_{\boldsymbol{l}}\delta_{j_Nk_N}\delta_{l_Nh_N}\\
	&= \sum_{\boldsymbol{s},\boldsymbol{t} \in \boldsymbol{\Omega}}e^{-i\boldsymbol{s}\boldsymbol{x}}e^{i\boldsymbol{t}\boldsymbol{x'}}c_{\boldsymbol{s}\boldsymbol{t}}
\end{align}

So we have demonstrated how the quantum kernel is computed, which include the small corrections to the equations presented in \cite{maria}.

\section{Comments on the cosine kernel example}

We now look at the example presented in \cite{maria} using equation (A5) from the paper to derive the cosine kernel. Despite the error in (A5), the correct cosine kernel is computed. We will now look closer into this example and conclude that in fact another error was made in the the computation that happened to cancel out the error in equation (A5). We also show how using \autoref{eq:full} gives the correct answer. We start with using \autoref{eq:full} with $N=1$ and $d=2$ and compare each step with the steps from \textbf{Example A.1.} in \cite{maria}.
\begin{align}
	k(\boldsymbol{x},\boldsymbol{x'}) &= \left|\sum_{j_1=1}^2\sum_{k_1=1}^2e^{-i(\lambda_{j_1}x-\lambda_{k_1}x')}V^*_{{k_1}1}V_{{j_1}1}\delta_{j_1,k_1}\right|^2 \\
	&= \left|\sum_{k_1=1}^2e^{-i\lambda_{k_1}(x-x')}V^*_{{k_1}1}V_{{k_1}1}\right|^2
\end{align}
We notice that we now have almost the same equation as in (A13). This means that from equation (A12) to (A13), the argument that was used must have been a worded version of requiring the $\delta_{j_1k_1}$ term. Let us now continue with the next step
 \begin{align}
	&= \left|e^{-i\lambda_1(x-x')}V^*_{11}V_{11}+e^{-i\lambda_2(x-x')}V^*_{21}V_{21}\right|^2\\
\end{align}
We now see that we have an expression similar to that of Equation (A14) in the paper with just one different term: $V^*_{21}$. Let us now examine what happens if we replace all the values with their numbers apart from that one different term:	
\begin{align}
	&= \left|e^{-i\left(-\frac{1}{2}\right)(x-x')}\frac{1}{\sqrt{2}}\frac{1}{\sqrt{2}}+e^{-i{\frac{1}{2}}(x-x')}V^*_{21}\left(-\frac{1}{\sqrt{2}}\right)\right|^2 \\
	&= \left|\frac{1}{2}e^{i\frac{x-x'}{2}}+V^*_{21}\left(-\frac{1}{\sqrt{2}}\right)e^{-i\frac{x-x'}{2}}\right|^2
\end{align}
If we continue this equation by replacing $V^*_{21}$ with $-\frac{1}{\sqrt{2}}$, we will reach equation (A16) in \cite{maria}. However, if we use $V^*_{12}$ as indicated in the paper, we will get the wrong sign and therefore the wrong answer. So this was the mistake which inadvertently corrected the previous mistake from Equation (A5). So for completeness sake we finish this derivation by using the correct replacement:
 \begin{align}
	&= \left|\frac{1}{2}e^{i\frac{x-x'}{2}}+\frac{1}{2}e^{-i\frac{x-x'}{2}}\right|^2 \\
	&= \left|\cos\left(\frac{x-x'}{2}\right)\right|^2 =\cos^2\left(\frac{x-x'}{2}\right)
\end{align}

\section{Acknowledgments}
This note was largely completed in 2023. Only the decision to post it, along with minor
additions such as this acknowledgments section, came later.
The author would like to thank B. David Clader for the several discussions in 2023 on this comment, and would like to thank Maria Schuld for her time in 2023 to briefly discuss this comment via an email.

\bibliographystyle{unsrt}
\bibliography{bibliography}

\end{document}